\documentclass[aps,nofootinbib,notitlepage]{revtex4}
\usepackage{graphicx}
\usepackage{xcolor}
\usepackage{stmaryrd}

\usepackage{amstext}
\usepackage{etoolbox}

\usepackage{makeidx}
\usepackage{amsfonts}
\usepackage{amsmath,amssymb,epsfig}
\usepackage{amscd}
\usepackage{amsthm}
\usepackage{mathrsfs}
\usepackage{dsfont}
\usepackage[applemac]{inputenc}
\usepackage[english]{babel}
\usepackage{enumitem} 
\usepackage{subfigure}
\usepackage{hyperref}
\usepackage{blindtext}
\usepackage{verbatim}
\usepackage{cancel}
\usepackage{cases}
\usepackage{tensor}

\definecolor{crimson}{rgb}{0.7, 0.08, 0.24}

\newcommand{\be}{\begin{equation}}

\newcommand{\ee}{\end{equation}}
\def\beqa{\begin{eqnarray}}
\def\eeqa{\end{eqnarray}}
\def\bean{\begin{eqnarray*}}
\def\eean{\end{eqnarray*}}
\def\nn{\nonumber}
\renewenvironment{thebibliography}[1]
         {\section*{References}\frenchspacing\small
          \begin{list}{[\arabic{enumi}]}
         {\usecounter{enumi}\parsep=2pt\topsep 0pt
         \settowidth{\labelwidth}{[#1]}
         \leftmargin=\labelwidth\advance\leftmargin\labelsep
         \rightmargin=0pt\itemsep=1pt\sloppy}}{\end{list}}

 \numberwithin{equation}{section}
 
\usepackage[normalem]{ulem}

\def\la{\langle}
\def\ra{\rangle}
\newcommand{\bra}[1]{\la {#1}|}
\newcommand{\ket}[1]{|{#1}\ra}

\newcommand{\R}{\mathbb{R}}

   \renewcommand{\t}{\tau}

\usepackage{graphicx}
\usepackage{epstopdf}
\usepackage{amsthm}
\usepackage{tocvsec2}
\usepackage{enumerate}
\usepackage[T1]{fontenc}
\usepackage{color}
\usepackage{xcolor}
\usepackage{mathtools}
\usepackage{bbm}
\usepackage{braket}
\usepackage{tcolorbox}
\usepackage{natbib}
\usepackage{graphicx}

\usepackage{pgfplots}

\pgfplotsset{compat = newest}

\definecolor{darkgreen}{rgb}{0.0, 0.2, 0.13}
\definecolor{darkred}{rgb}{0.55, 0.0, 0.0}
\definecolor{carrotorange}{rgb}{0.93, 0.57, 0.13}
\usepackage{hyperref}
\hypersetup{
	colorlinks=true,
	linkcolor=blue,
	filecolor=magenta,
	urlcolor=cyan,
}
\usepackage[squaren,Gray]{SIunits}

\newcommand{\va}{\scriptscriptstyle}
\newcommand{\van}{\scriptstyle}

\newcommand{\bee}{\nopagebreak[3]\begin{equation*}}
\newcommand{\eee}{\end{equation*}}

\newcommand{\ba}{\nopagebreak[3]\begin{eqnarray}}
\newcommand{\ea}{\end{eqnarray}}
\DeclareFontFamily{U}{rsfs}{}         
\DeclareFontShape{U}{rsfs}{m}{n}{<5> rsfs5 <6><7> rsfs7          %
  <8><9><10><10.95><12><14.4><17.28><20.74><24.88> rsfs10}{}     %
\DeclareMathAlphabet{\mathfs}{U}{rsfs}{m}{n}                     %
\newcommand{\mfs}[1]{\mathfs {#1}}                               %

\newcommand{\sH}{{\mfs H}}
\newcommand{\sL}{{\mfs L}}

\newcommand{\sI}{{\mfs I}}
\newcommand{\sO}{{\mfs O}}

 \usepackage{braket}

\usepackage{etoolbox}

\begin{document}
\title{Modelling quantum particles falling into a black hole: the deep interior limit}

\author{Alejandro Perez}
\affiliation{{Aix Marseille Universit\'e, Universit\'e de Toulon, CNRS, CPT, Marseille, France}}
\date{\today}

\author{Salvatore Ribisi}
\affiliation{{Aix Marseille Universit\'e, Universit\'e de Toulon, CNRS, CPT, Marseille, France}}

\author{Sami Viollet}
\affiliation{{Aix Marseille Universit\'e, Universit\'e de Toulon, CNRS, CPT, Marseille, France}}

\begin{abstract}
In this paper we construct a solvable toy model of the quantum dynamics of the interior of a spherical black hole with falling spherical scalar field excitations.
We first argue about how some aspects of the quantum gravity dynamics of realistic black holes emitting Hawking radiation can be modelled using Kantowski-Sachs solutions with a massless scalar field when one focuses on the deep interior region $r\ll M$ (including the singularity). Further, we show that in the $r\ll M$ regime, and in suitable variables, the KS model becomes exactly solvable at both the classical and quantum levels. The quantum dynamics inspired by loop quantum gravity is revisited. We propose a natural polymer-quantization where the area $a$ of the orbits of the rotation group is quantized. The polymer (or loop) dynamics is closely related with the Schroedinger dynamics away from the singularity with a form of continuum limit naturally emerging from the polymer treatment. The Dirac observable associated to the mass is quantized and shown to have an infinite degeneracy associated to the so-called $\epsilon$-sectors. Suitable continuum superpositions of these are well defined distributions in the fundamental Hilbert space and satisfy the continuum Schroedinger dynamics.
\end{abstract}

\maketitle

\section{Motivation}

The fate of the singularities of general relativity is a central question for quantum gravity  that  concerns important physical situations such as those arising in (big-bang) cosmologies and  black hole formation and evaporation.  One of the central features of loop quantum gravity  is the inherent discreteness of quantum geometry at the Planck scale. The lack of smoothness of the geometry at the fundamental level challenges the classical view of the singularities of general relativity as a frontier of spacetime geometry, and  strongly suggests the possibility of  a microscopic dynamical description that could define dynamics beyond the limit where classical description fails. 

The history of the approach starts with the discovery of Ashtekar's connection variables which first suggested that the quantum dynamical evolution equations of gravity might admit a background independent finite and non perturbative formulation \cite{Ashtekar:1986yd, Ashtekar:1987gu}. 
This initial suggestion grew into the approach of loop quantum gravity (LQG) with the contribution of many (for reviews and text books see \cite{Perez:2004hj, Ashtekar:2004eh, rovelli_2004, thiemann_2007,  Agullo:2016tjh}). The LQG approach has produced insights about the possible nature of matter and geometry at the Planck scale and has led to new ideas about the origin of black hole entropy, the generation of quantum effects in early cosmology, and stimulated hopes about the possible regularizing role of Planckian granularity (for quantum field theory and gravity). However, a clear understanding of the question of the fate of singularities in realistic physical situations has remained a difficult one, as addressing it would actually require the (still lacking) complete dynamical control of LQG in  situations involving matter and geometry degrees of freedom in the deep ultraviolet regime in full generality.  

Nevertheless, the view that the evolution across singularities should be well behaved has become consensual in the field over time thanks to the accumulated experience in simple low dimensional as well as symmetry reduced models of black holes and cosmology.  Professor Abhay Ashtekar has been one of the key leading driving forces along this path, and main defender of the view (to which we adhere) that dynamics across the {\em would-be-singularity} should be well defined in the quantum theory. It is a pleasure to contribute to this special issue with this work that, we believe, is representative of this standpoint.

The first examples of singularity avoiding models where found in the context of quantum cosmology by Martin Bojowald \cite{Bojowald:2001xe}. This seminal work grew later into a large number of contributions in the field now known as loop quantum cosmology \cite{Bojowald:2005epg, Ashtekar:2003hd, Ashtekar:2011ni}. 
Even in these simple models the quantum dynamics can be rather involved. However, it was soon realized \cite{Taveras:2008ke} that effective semiclassical equations could be used to describe the dynamics across the singularity and that these equations were quite easy to describe.  The domain of applicability of these techniques was extended in a variety of manners to models involving black holes \cite{Modesto:2004xx, Ashtekar:2005qt, Modesto:2005zm, Gambini:2013ooa, Gambini:2008dy, Gambini:2013hna, Gambini:2014qga, Ashtekar:2018lag} (for reviews see \cite{Olmedo:2016ddn, Gambini:2022hxr, Ashtekar:2023cod}, for quantum modifications inspired by other approaches see \cite{Nicolini:2019irw, Koch:2014cqa, Saueressig:2015xua}).  In many of the latter cases the natural starting point has been to consider the quantum dynamics of the interior of spherically symmetric and static spacetimes of the Kantowski-Sachs type (the Schwarzschild black hole interior in the vacuum case). In all these cases the interior singularity is removed and replaced by a quantum transition across what would have been the singularity in the classical description realizing aspects of existing scenarios \cite{bardeen1968non, Hayward:2005gi, DeLorenzo:2014pta, Frolov:2016pav}.

Simple models are nice as they illustrate possibly generic features of the general situation. However, they carry the drawback of being often removed, by the very symmetry assumptions that simplify them, from the realistic physical situations about which one would like to gain non-trivial insights. Moreover, when it comes to black holes, most of the studies have focused on effective dynamical descriptions, while quantum dynamics has received less attention due to its often unsurmountable complexity even in simple models. For instance, concerning the first drawback we know that real black holes are not time translational invariant due to the expected presence of Hawking evaporation (in contrast with the static nature of many of the quantum black hole models) and that all symmetry assumptions must fail near the singularity when the back-reaction of Hawking particles correlated with the outside radiation would be properly taken into account (see  \cite{Perez:2022jlm} for further discussion of this issue). When it comes to the second drawback, even when effective descriptions can provide the dynamical evolution of the spacetime geometry with matter fields on it, its classical nature precludes the analysis of genuine quantum phenomena such as entanglement and other quantum information issues of highest interest from the perspective of Hawking's evaporation (e.g. the longstanding information puzzle or the question of the fate of unitarity in black hole evaporation). Even when our model will not resolve the first limitation, we believe that it provides a humble small step in the right direction. Concerning the second, we will see that the quantum dynamics is fully accessible in our simple model opening the road for exact calculations in the quantum realm. 
\begin{figure}[h!]
	{\includegraphics[height=9cm]{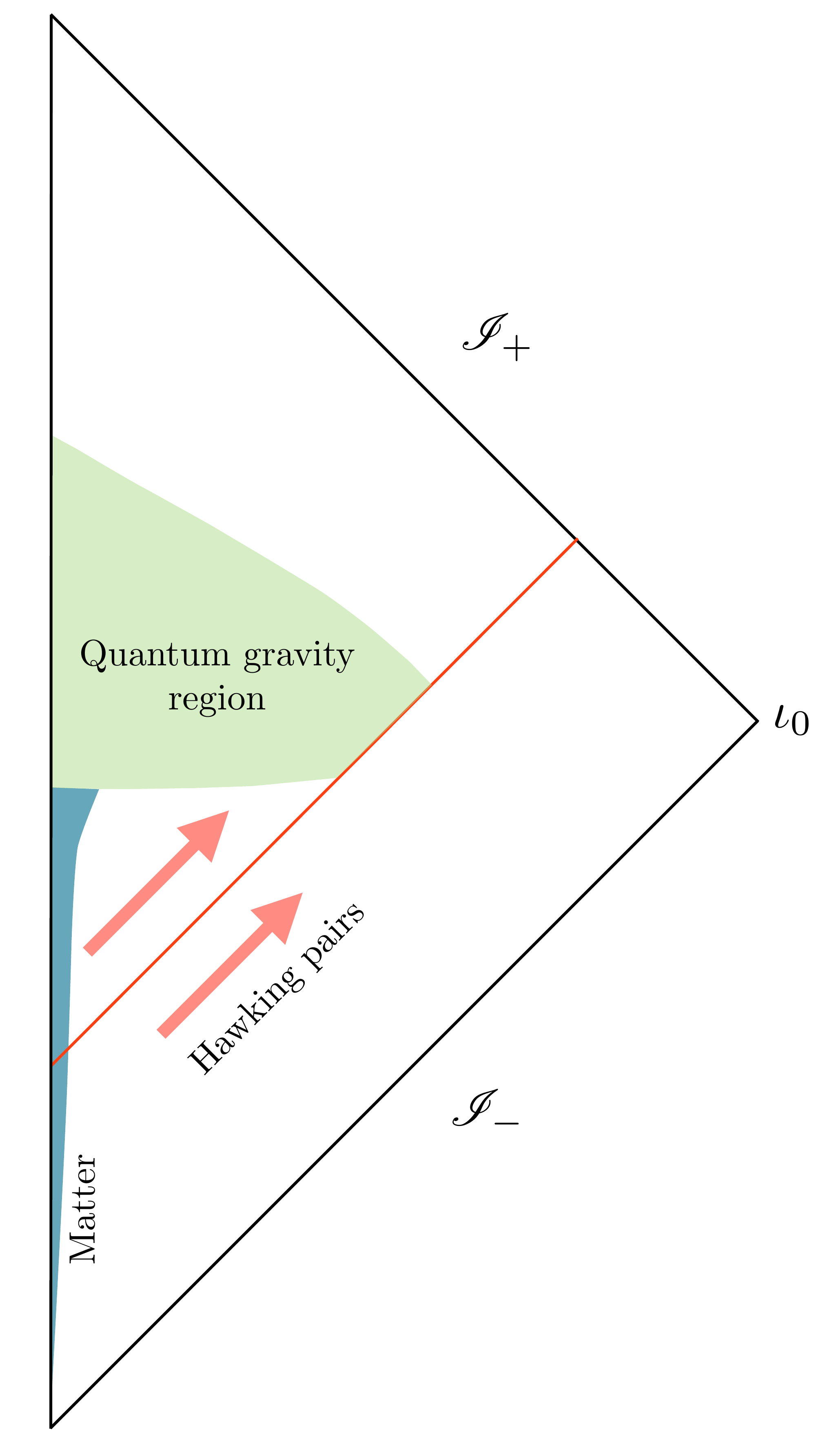}}
	\caption{The Ashtekar-Bojowald paradigm}.
	\label{AA-MB}
\end{figure}

The interior region $r< 2M$ of a Schwarzschild black hole of mass $M$ can be seen as a homogeneous anisotropic cosmological model where the $r$=constant surfaces (in the usual Schwarzschild coordinates) are Cauchy surfaces of homogeneity where any two arbitrary points can be connected along orbits of the isometry group that involves spacelike translations along the staticity Killing field $\xi=\partial_t$ and the rotations associated to spherical symmetry. Models with these isometries will be refereed to as  Kantowski-Sachs (KS) models \cite{kantowski1966some}. They include not only the Schwarzschild black hole interior geometry (vacuum case) but also the Reissner-Nordstrom black hole interior geometry (in the Einstein-Maxwell case) and other solutions depending of the type of matter that one decides to couple to the system.
In this paper we would like to emphasize the fact that Kantowski-Sachs models (with a massless scalar field coupling) define a natural toy model capturing some (possibly interesting) aspects of the dynamics and back-reaction of matter near the singularity of realistic black holes that Hawking radiate and evaporate. The model can certainly not replace the full dynamical description  of a generic gravitational collapse in the full theory as it remains a toy model with finitely many degrees of freedom. However, we will argue, it can handle in a simplistic way some dynamical aspects that might be relevant when discussing questions  in the context of evaporating black holes.

\subsection{Scalar excitations falling inside a Schwarzschild black hole: the deep interior regime}\label{esto}

Let us consider a free test point particle (with no angular momentum) falling into the interior of a Schwarzschild black hole.
As the particle approaches the singularity---in a description on an $r$ equal constant slicing of the interior---one expects its wave function to become better and better approximated by a translational invariant wave function since the expansion in the spacelike Killing direction $\xi=\partial_t$ diverges for $r\to 0$. If this conclusion is correct then it means that zero angular momentum test particles can be approximated by the type of excitations that can be accommodated in the dynamical framework of the KS cosmologies (at least in the sense of a near singularity approximation).  One can be quantitative about this intuition as follows:
 free test particles with four wave vector $k^a$ on the Schwarzschild background are associated with the conserved Killing energy ${\cal E}\equiv -k^a\xi_a$. We are assuming that the particle has zero angular momentum which implies that its wave function is already translational invariant in the directions transversal to $\xi^a$ on the $r$-slices. The wave function can only vary in the direction of the Killing $\xi^a$ and  the component of the physical momentum in this direction is given by 
 \be\label{lapu}
 p_\xi \equiv \frac{k^a\xi_a}{\sqrt{\xi\cdot\xi}}=-{\cal E}\sqrt{\frac{r}{2M-r}},
 \ee
 which vanishes in the limit $r\to 0$. The wave length of such a particle diverges, and thus particles without angular momentum are better and better represented by translational invariant excitations as one approaches the singularity. These are precisely the kind of homogeneous configurations that can be described in the KS framework. 
 
This simple implication deduced from the idealized notion of test particle can be made more precise by looking at the analogous features of scalar field excitations (solutions of the Klein-Gordon equation). Indeed, the simplistic argument given here can be made precise in the field theoretical context as we will show in what follows.

\subsubsection{Solutions of the Klein-Gordon equation in the deep interior region}
Here we argue that the Kantowski-Sachs model (described in detail in Section \ref{KSM}) coupled to a massless scalar field
faithfully captures the dynamics of a Klein-Gordon excitation falling into the deep interior region of a Schwarzschild black-hole. We will do this by analysing the behaviour of solutions of the Klein-Gordon equation on the Schwarzschild background  in the $r \ll 2M$ regime. We will focus on the spherically symmetric solutions and  show that they become homogeneous on $r$=constant surfaces as $r\to 0$ and thus can be accommodated in the framework of KS configurations.
This implies that the KS system can be used to model the dynamics and (most importantly) the back-reaction of such (zero angular momentum) scalar configurations falling into the deep interior region of spherically symmetric black holes.

Let us first start by approximating the Schwarzschild metric in the deep interior region  $r \ll 2M$ as \begin{align}
	d s^2 = \frac{2M}{r} d t^2 - \frac{r}{2M} d r^2 + r^2 d \Omega^2 .
\end{align}
We will choose coordinates such that the time-radial part of the metric is conformally flat. Remarkably, in the deep interior region,  this is achieved by switching to area variables $a= 4\pi r^2$ (the well known tortoise coordinate $r_*$ is actually proportional to $r^2$ near the singularity). With these variables the metric becomes
\begin{align}
	ds^2 
	&=  \frac{1}{16 \pi \sqrt{a_{\rm H} a}} \left( d \t^2 - d a^2  \right) + \frac{a}{4\pi} d \Omega^2 ,
\end{align}
with  $a_{\rm H}= 16\pi M^2$, $\tau= \sqrt{16 \pi a_{\rm H}}  t$, and $a=4\pi r^2$.
The Klein-Gordon equation for a massless scalar field then reads
\begin{align}
	\square \Phi &= \frac{1}{\sqrt{-g}} \partial_\mu \left( \sqrt{-g} g^{\mu\nu} \partial_\nu \Phi \right)=0  \\
	&\Longleftrightarrow \left( - \frac{\partial^2 \Phi}{\partial a^2} - \frac{1}{a} \frac{\partial \Phi}{\partial a} + \frac{\partial^2 \Phi}{\partial \tau^2} +  \frac{1}{4 \sqrt{a_{\rm H} a} a} \left( \frac{1}{\sin\theta} \frac{\partial \left( \sin\theta \frac{\partial \Phi}{\partial \theta} \right)}{\partial \theta} + \frac{1}{\sin^2 \theta} \frac{\partial^2 \Phi}{\partial \varphi^2} \right) \right) = 0 .
\end{align}
To solve it, we make the usual ansatz
\begin{align}
	\Phi_{\ell m} = e^{i \omega t} Y_{lm} (\theta, \varphi) \phi_l (a) = e^{i\frac{\omega \tau}{\sqrt{16\pi a_{\rm H}}}} Y_{lm} (\theta, \varphi) \phi_l (a)
\end{align}
which reduces the Klein-Gordon equation to
\begin{align}\label{kgedi}
	 \phi''_l + \frac{\phi'_l}{a} + \left( \frac{\omega^2}{16\pi a_{\rm H}} + \frac{l (l+1)}{4 \sqrt{a_{\rm H} a} a}  \right) \phi_l = 0 .
\end{align}
For the spherical modes $l=0$ one obtains
\begin{align}
	\phi_0(a) = c_1 J_0\left(\frac{\omega a}{\sqrt{16\pi a_{\rm H}}}\right)+c_2 Y_0\left(\frac{\omega a}{\sqrt{16\pi a_{\rm H}}}\right),
\end{align}
with $J_0$ and $Y_0$ Bessel functions and $c_{1}$ and $c_2$ constants. 
As we will prove in Section \ref{KSM} that these solutions match nicely with solutions of the KS model (mentioned at the end of the introduction).
This follows from two complementary properties of the solutions of the Klein-Gordon equation. On the one hand, the near singularity limit of the quantity (simply related to the KS momentum variable as we will see in Section \ref{KSM})
\be
\lim_{a \rightarrow 0}  a \frac{\partial \phi_{0}}{\partial a}= -\frac{2c_2}{\pi} 
\ee 
is finite and independent of $\omega$. On the other hand, the $t$ dependence of the Klein-Gordon solutions is `ironed' by the infinite expansion of the geometry 
in the $\partial_t$ direction: the region $\ell_0\equiv \Delta t=2\pi \omega^{-1}$ where the solution has a significant (order-one) change corresponds to a length scale $\Delta d\approx2\pi  \omega^{-1} \sqrt{M/r}$ (in agreement with the infinite redshift effect captured in equation \eqref{lapu}).  Therefore, in a length scale $\ell_p \ll \ell \ll \Delta d$ along the background Killing field direction $\xi=\partial_t$, and in the deep interior region $a\ll M^2$, the solutions of the Klein-Gordon equations can be considered as homogeneous and therefore compatible with initial data that would be admissible in the KS model.  

We will see in Section \ref{KSM} that the momentum variable ${\rm p}_{\phi}$ in the KS model is simply related to the quantity whose limit was considered in the previous paragraph as  
it is defined as
\begin{align}
	{\rm p}_{\phi} &\equiv -8 \pi  M \ell_0 r \frac{\partial \Phi_{00}}{\partial r} = -\ell_0 a_{\rm H}a \frac{\partial \Phi_{00}}{\partial a},
 \end{align}
where the pre-factors arise form the hamiltonian analysis of Section \ref{KSM} and $\ell_0$ is an IR cutoff scale naturally associated in the previous discussion to 
the scale $\Delta t=2\pi \omega^{-1}$. It follows from the previous considerations that
\begin{align}\label{radio}
	\lim_{a \rightarrow 0} {\rm p}_{\phi} = {\rm constant}
\end{align}
in the region of interest.
One can relate the previous quantity to the average `energy density' on the $r$=constant  hyper-surfaces as one approaches the singularity  
(this will be simply related to the KS Hamiltonian that will be defined in the following section). Namely,
\be \frac{1}{\ell_0} \int_{t_0}^{t_0+\ell_0} \!\!dt\left(\int d\theta d\varphi \left(\sqrt{|g|} T_{\mu\nu} \partial_r^\mu\partial_r^\nu \right)  \right)
=\frac{{\rm p}_{\phi}^2}{64 \pi M^2 \ell^2_0},\ee
where the scale $\ell_0$ enters the definition of the average in the time direction. For concreteness one can match $\ell_0$ to the wavelength $2\pi \omega^{-1}$ of the excitation and the previous result will already hold (of course it holds for $\ell_0>2\pi \omega^{-1}$). 

For completeness we give the limiting behaviour of solutions in the non-spherical case. For the non-spherical modes, one can neglect the term containing the frequency in equation \eqref{kgedi} in analysing the small $a$ behaviour of solutions. If we do so, we obtain for $l \ne 0$
\begin{align}
	\phi_l (a)&= c_1 J_0\left(\sqrt[4]{\frac{a}{\pi}} \sqrt{\frac{l (l+1)}{M}}\right)+2 c_2 Y_0\left(\sqrt[4]{\frac{a}{\pi}} \sqrt{\frac{l (l+1)}{M}}\right).
\end{align}
Solutions diverge logarithmically (as $\log[a]$) for $a\to 0$. This holds both for spherically symmetric as well as for non spherically symmetric solutions as it follows from the asymptotic behaviour of the Bessel functions or from the finiteness of ${\rm p}_{\phi}$ in the spherical case. The mild character of the divergence was emphasized in \cite{Ashtekar:2022cih, Ashtekar:2022oyq} as an attractive possibly interesting property  when one considers the definition of the associated quantum operators in quantum field theory (in view of a possible definition of semiclassical gravity). Here we simply point out that such simple behaviour allows for bridging to a solvable KS model to understand aspects of the back-reaction of classical (as well as quantum) excitations falling into a spherically symmetric black hole. 

%
%
\section{The Kantowski-Sachs spacetime coupled with a massless scalar field}\label{KSM}

In this section we revisit the construction of the phase space of the KS model by perforing the canonical analysis of the associated symmetry reduced model where staticity and spherical symmetry are imposed from the onset (in Section \ref{ps}). To improve the clarity of the presentation we simply start from the vacuum case---whose solutions are isomorphic to the interior Schwarzschild solutions---and later couple the system to a scalar field without mass. We express variables in terms of the usual Schwarzschild-like coordinates. In Section \ref{did} we present a truncation of the Hamiltonian and show, in Section \ref{solubili}, that it defines a tractable approximation of the dynamics in the $r\ll M$ region of the interior of physically realistic black holes.
We call this regime the {\em deep interior dynamics}. In Section \ref{vhr} we show that the regime of applicability of the model includes the physically interesting situation of Hawking scalar excitations with zero angular momentum falling inside of the black hole.  

\subsection{Symmetry reduced covariant phase space} \label{ps}

It is well known that for a spherically symmetric and static spacetime, the line element can be written without any loss of generality as 
\begin{equation}\label{ansatz}
ds^2=-f(r)dt^2+h(r)dr^2+r^2d\Omega^2 \ .
\end{equation}
It follows that the Eintein-Hilbert action (with the appropriate boundary term that renders it differentiable) becomes
\be\label{ss}
S_{\rm geo}=\frac{1}{16\pi }\left[\int_R d^4x\sqrt{-g}R + 2 \int_{\partial R} K\right] =\frac{\ell_0}{2\ell_p^2} \int dr \left(\sqrt{fh}+\sqrt{\frac{f}{h}} +\frac{\dot f r}{\sqrt{fh}}\right),\ee
where the dot denotes the derivative with respect to $r$ and $\ell_0$ is a cut-off in the non compact spacelike $\partial_t$ direction that regularizes the dynamical system. The cut-off will be associated a natural meaning in modelling the fate of zero angular momentum excitations falling inside the black hole. In the {deep interior region} $r\ll M$ and we will take $\ell_0\sim \omega^{-1}$ for $\omega\approx 1/M$ (the typical frequency in the Hawking spectrum of a macroscopic black hole of mass $M$). 
One can easily verify that the variations of the action lead to the Schwarzschild solutions
\be\label{schww}
ds^2=-{\rm p}_M^2\left(1-\frac{2M}{r}\right)dt^2+\frac{dr^2}{\left(1-\frac{2M}{r}\right)}+r^2 d\Omega^2,
\ee
and the symplectic potential (stemming from the on-shell evaluation of the action variation)
\be
\theta=-\frac{\ell_0}{\ell_p^2} 
   (c_1 {dM}+2
   {M} {d{\rm p}_M}-2
   {d{\rm p}_M} r),
\ee
and symplectic structure
\be
\omega=\frac{\ell_0}{\ell_p^2}
   d{\rm p}_M \wedge {dM}.
\ee
Instead of working directly with the physical phase space parametrized by the Dirac observables ${\rm p}_M$ and $M$ it will be convenient for us to work with kinematical variables and constraints for the moment. This is because of the usual difficulty in linking  the timeless physical phase space with a classical intuition based on spacetime geometry. To avoid such difficulties we would like to have a notion of parametrized `time evolution' which in our context will take the form of an area radius evolution.  Thus we take the integrand of \eqref{ss} as the Lagrangian $\sL_{\textrm{\rm geo}}$ of the spacetime subsystem
\begin{equation}
\sL_{\rm geo}=\frac{\ell_0}{2\ell_p^2} \left(\sqrt{fh}+\sqrt{\frac{f}{h}}+\frac{r\dot f}{\sqrt{fh}}\right) .
\end{equation}
On the other hand, we will couple the system to a massless scalar field by adding the matter action
\be
S_m=-\frac{1}{2}\int_R d^4x\sqrt{-g}\partial_a\phi \partial^a\phi =-2\pi \ell_0 \int dr r^2 \dot \phi ^2\sqrt{\frac{f}{h}}.
\ee
The conjugate momenta to $f,h$ and $\phi$ are given by
\begin{equation}\label{conj m}
{\rm p}_f=\frac{\ell_0}{2\ell_p^2}\frac{r}{\sqrt{fh}} \quad , \quad {\rm p}_h=0 \quad \textrm{and} \quad {\rm p}_\phi=-4\pi r^2 \ell_0 \sqrt{\frac{f}{h}} \dot \phi,
\end{equation}
and the primary Hamiltonian, defined by $H=\dot f {\rm p}_f+\dot h {\rm p}_h+\dot \phi {\rm p}_\phi-\sL_{\phi}-\sL_{\rm geo}$, becomes
\begin{equation} \label{p H}
{\rm H}_1=-\frac{\ell_0}{2\ell_p^2} \frac{f(h+1)}{\sqrt{fh}}
-\frac{h{\rm p}_\phi^2}{8\pi r^2 \ell_0 \sqrt{fh}} \ .
\end{equation}
From the expression of the conjugate momenta \eqref{conj m}  we identify the constraints \begin{equation}\label{constraints}
\xi\equiv {\rm p}_f-\frac{\ell_0}{2\ell_p^2} \frac{r}{\sqrt{fh}}=0 \quad \textrm{and} \quad {\rm p}_h=0,
\end{equation}
and the secondary Hamiltonian 
\begin{equation}\label{s H}
	{\rm H}_2={\rm H}_1+\lambda \xi + \eta {\rm p}_h \ ,
\end{equation}
where $\lambda$ and $\eta$ are Lagrange multipliers. One can show that the stability of the two constraints \eqref{constraints} can be ensured by fixing  the associated Lagrange multipliers, i.e.,  the constraints \eqref{constraints} are second class and can be explicitly solved leading to  
\begin{equation}\label{central}
	{\rm p}_h=0 \quad \textrm{and} \quad h= \frac{\ell^2_0}{4\ell_p^4}\frac{r^2}{f{\rm p}_f^2} .
\end{equation}
Thus, the secondary Hamiltonian \eqref{s H} reduces to
\begin{equation}\label{wewe}
	{\rm H}_2=-\frac{1}{r}\left(f {\rm p}_f+\frac{1}{16\pi \ell_p^2} \frac{{\rm p}_{\phi}^2}{f {\rm p}_f}+\frac{\ell^2_0}{4\ell_p^4} \frac{r^2}{{\rm p}_f} \right)\ .
\end{equation}
The previous encodes the KS dynamics of geometry coupled to a massless scalar field.
The relevant solutions for physical applications correspond to small departures from the vacuum Schwarzschild solutions representing macroscopic black holes with scalar field perturbation falling inside. We will further simplify the system by focusing on, what we call, the 
{deep interior region} $r\ll M$ where $M$ is the mass scale defined by the corresponding black hole solution perturbed (in the sense of Sections \ref{solubili} and Section \ref{vhr}) by the presence of matter. It is in this regime where the solutions of the KS system faithfully describe the dynamics of a spherically symmetric scalar perturbation (representing for instance a Hawking particle) as it falls towards the interior singularity.  The KS Hamiltonian evolution matches, in this sense, the test-field evolution (the Klein-Gordon solutions on the Schwarzschild background fixed non dynamical background) and incorporates, as a simplified model, aspects of the back-reaction that are expected to become more important as one approaches the singularity.

\subsection{The deep interior dynamics}\label{did}

We are interested in the dynamical evolution in  the  $r  \ll 2M$ regime. In addition we will use the present dynamical system to model (in a suitable approximation) the back-reaction  of a Hawking quantum falling into a black hole singularity. Hawking particles  do not correspond to static excitations as the one we can model with the symmetry assumptions of the present section. However, as argued in Section \ref{esto}, when spherically symmetric, these particles look more and more static as seen by a radially freely falling observer in the limit $r\to 0$.
This is the reason why we are interested in such regime of the present dynamical system. In the next section we will study the classical solution of the model using perturbation theory in the parameter ${\rm p} _{\phi}^2/M^2$---as ${\rm p} _{\phi}^2$ will be assumed to be much smaller to $M^2$ in applications---and show that the dynamics simplifies in the deep interior region. 
The simplification occurs due to the negligible effect of the last term in the expression of the hamiltonian \eqref{wewe}: more precisely, in the deep interior region, the  Hamiltonian is well approximated by  
\begin{equation}\label{HDI}
	{\rm H}_{\rm di}=-\frac{1}{r}\left(f {\rm p}_f+\frac{1}{16\pi \ell_p^2} \frac{{\rm p}_{\phi}^2}{f {\rm p}_f}\right).
\end{equation}
This toy theory reflects the dynamics of the leading order in an expansion near $r=0$. Order $\sO(r)$ effects could be included in perturbation theory near $r=0$ in which case the term we dropped would correspond to the perturbation Hamiltonian. The consistency of this truncation will be shown in Section \ref{solubili}.

The system we are dealing with has no gauge symmetries as the radial reparametrization symmetry  has  been gauged fixed with the metric ansatz \eqref{ansatz} by choosing the area radius as time. In order to recover the structure of a gauge theory, with a clear analogy with the full theory of LQG, it will be convenient `reparametrize' the system by promoting the area radius $r$ to a degree of freedom with conjugate momentum ${\rm p}_r$ and add a scalar constraint ${\rm C}={\rm p}_r-{\rm H}_2=0$. The  phase space is therefore extended to $(f,{\rm p}_f,r,{\rm p}_r,\phi,{\rm p}_\phi)$, and the number of degrees of freedom is preserved by the inclusion of the Hamiltonian constraint 
\begin{equation}\label{Hc 1}
{\rm C}_r={\rm p}_r-\frac{1}{r}\left(f {\rm p}_f+\frac{{\rm p}_{\phi}^2}{16\pi \ell_p^2 f {\rm p}_f}\right) \approx 0 .
\end{equation} 
In this approximation one can show that we have the following Dirac observables
\begin{equation}
{\rm D}_1=f{\rm p}_f \quad , \quad {\rm D}_2=fr^{-\frac{{\rm p}_\phi^2}{4f {\rm p}_f^2}+1} \quad , \quad {\rm D}_3={\rm p}_\phi \quad , \textrm{and} \quad {\rm D}_4=\phi+\frac{{\rm p}_\phi\log(r)}{2f{\rm p}_f} .
\end{equation}
It will be convenient to make the following canonical transformation and thus introduce 
what we call the {\em deep interior variables} 
\begin{equation}\label{nsv}
	m=-f {\rm p}_f \ \ \ \textrm{and} \ \ \ {\rm p}_m=-\log(-f) \ ,
\end{equation}
and---in trying to introduce the kinematical structure proper to loop quantum gravity---to adopt the area $a$ of the surfaces of constant $r$, namely 
\begin{equation}
	a=4\pi r^2 \quad \textrm{and} \quad {\rm p}_a=\frac{{\rm p}_r}{8\pi r} \ ,
\end{equation}
as new dynamical variables.
With this choice the phase space is described by the geometric variables $m,{\rm p}_m,a$ and ${\rm p}_a$ with Poisson brackets
\begin{align*} \label{classical ccr}
	\{m,{\rm p}_m\}&=1 \ , \quad \{a,{\rm p}_a\}=1
\ ,
\end{align*}
and by the matter variables $\phi$ and ${\rm p}_\phi$ for which
\begin{equation}\label{classical ccr mat}
	\{\phi,{\rm p}_\phi\}=1,
\end{equation}
with all the other Poisson brackets equal to zero.
In the new variables the deep interior dynamics Hamiltonian constraint \eqref{Hc 1} becomes 
\begin{equation} \label{Hc 2}
\boxed{{\rm C}_a={\rm p}_a+\frac{1}{2a}\left(m+\frac{{\rm p}_{\phi}^2}{16\pi \ell_p^2 m}\right) \approx 0 .}
\end{equation}
The previous constraint is central in the rest of the paper. We will see that it leads to a fully controllable dynamics both at the classical as well as the quantum level.
Indeed, the dynamics is exactly solvable in the vacuum case while it can be dealt with in perturbation theory for the case where the scalar field is excited. 
In the next section we will justify the truncation that took us from \eqref{wewe} to \eqref{HDI}  (and finally to the constraint \eqref{Hc 2}) using perturbation theory. In Section \ref{vhr} we will show that the perturbative regime is consistent with the conditions that make our model applicable to the description of a spherically symmetry Hawking particle falling into a Schwarzschild black hole during evaporation. 

\subsection{Perturbative solutions in ${\rm p}_{\phi}/M$ and the dynamics in the deep interior region}\label{solubili}

Exact KS solutions with scalar fields have been studied in the past (see for instance \cite{doi:10.1063/1.529717}). KS coupled to scalar fields does not lead necessarily to (asymptotically flat) back hole spacetimes globally speaking. However,  we will show here that solutions can be interpreted in terms of perturbations of a vacuum Schwarzschild solution in the deep interior region $r\ll 2M$ in the regime where ${\rm p}_{\phi}/M\ll 1$.  We will also show that in that regime the Hamiltonian \eqref{wewe}, and the equations it generates, can be well approximated by \eqref{HDI}. This will lead to a simple solvable system, both at the classical and quantum levels, which can be used to model aspects of the physics of (zero angular momentum) scalar particles falling into a spherically symmetric black hole (possibly useful in view of describing aspects of Hawking radiation). We will analyze the system in first order perturbation theory in ${\rm p}_{\phi}/M$. 

In order to best organise the perturbative equations we replace ${\rm p}_\phi^2$ by $\epsilon^2 {\rm p}_\phi^2$ where $\epsilon$ is a smallness parameter. We introduce the following expansion of the relevant dynamical quantities 
\begin{align}
	&f(r)=f_0(r)+\epsilon^2f_1(r)+\sO(\epsilon^4),\\
	&{\rm p}_f(r)={\rm p}_{f0}(r)+\epsilon^2{\rm p}_{f1}(r)+\sO(\epsilon^4) \ ,
\end{align}
and write the equations of motion for them by keeping terms up to order $\epsilon^2$. Starting from \begin{align} \label{eq f pf}
	&\dot f=\{f,{\rm H}_2\}=-\frac{f(r)}{r} +{\frac{\epsilon^2{\rm p}_\phi^2}{16 \pi \ell_p^2 r f(r){\rm p}_f(r)^2}+\frac{\ell_0^2r}{4\ell_p^4 {\rm p}_f(r)^2}}, \nn \\
	&\dot {\rm p}_f=\{{\rm p}_f ,{\rm H}_2\}={\frac{{\rm p}_f(r)}{r}-\frac{\epsilon^2{\rm p}_\phi^2}{16 \pi \ell_p^2 r f(r)^2{\rm p}_f(r)}},
\end{align}
with ${\rm H}_2$ given in equation \eqref{wewe}, one can solve the equations order by order 
with solutions
\begin{align} \label{fpf order 0}
	&f_0(r)={\rm p}_M^2\left(1-\frac{2M}{r}\right) , &f_1(r)=-\frac{\ell_p^2{\rm p}_\phi^2}{8\pi r\ell_0^2M^2} \left(2M+(M-r)\log\left(\frac{r}{2M-r}\right)\right) , \\
	&{\rm p}_{f0}(r)=\frac{\ell_0}{4\ell_p^2{\rm p}_M}r , &{\rm p}_{f1}(r)=-\frac{{\rm p}_\phi^2r}{32\pi\ell_0 M^2{\rm p}_M^3} \left(\frac{2M}{2M-r}+\log\left(\frac{r}{2M-r}\right)\right).
\end{align}
The function $h(r)$, is recovered from the constraint \eqref{constraints} and gives
\begin{equation}\label{h order 1}
h_0(r)=\frac{1}{1-\frac{2M}{r}},\ \ \ 
	h_1(r)=-\frac{\ell_p^2{\rm p}_\phi^2r\log\left(\frac{r}{2M-r}\right)}{8\pi\ell_0^2{\rm p}_M^2M(2M-r)^2} \ .
\end{equation}
Note that the Schwarzschild solution with mass $M$ and conjugate momentum ${\rm p}_M$---as in \eqref{schww}---is recovered in the leading order.
Consistency of the perturbative treatment requires the ratio of first-to-leading order contributions to remain small, namely
\begin{align}\label{validity}
	&\frac{f_1(r)}{f_0(r)} =\frac{\ell_p^2{\rm p}_\phi^2}{32\pi \ell_0^2 M^2 {\rm p}_M^2}\log\left(\frac{r^2}{4M^2}\right)+\sO\left(\frac rM\right) \ll1, \nn \\
	&\frac{h_1(r)}{h_0(r)} = \frac{\ell_p^2{\rm p}_\phi^2}{32\pi \ell_0^2 M^2 {\rm p}_M^2}\log\left(\frac{r^2}{4M^2}\right)+\sO\left(\frac rM\right) \ll1 
\end{align}
in the regime $r<M$.  The logarithmic behaviour of the right hand side of the previous equations is not a threat, as the classical solutions are not to be trusted for $r$ near the Planck scale. Therefore, assuming $r>\ell_p$, we conclude that our analysis is consistent as long as  
\begin{equation}\label{validity 2}
\frac{\ell_p^2{\rm p}_\phi^2}{32\pi \ell_0^2 M^2 {\rm p}_M^2}\log\left(\frac{\ell_p^2}{4M^2}\right) \ll 1
\end{equation}
which is always valid in the usual  macroscopic black hole regime $\ell_p/M\ll 1$. Similarly, note that 
the ratio of the last term in the Hamiltonian \eqref{wewe} to the leading first term is exactly $h(r)$ which in the same regime is $\sO(\ell^2_p/M^2 \log (\ell_p/M))$ as seen from \eqref{h order 1}. This suggests that one can simplify the dynamics (if interested in the deep interior region $r\ll M$) and use the Hamiltonian
\begin{equation}
	{\rm H}_{\rm di}=-\frac{1}{r}\left(f {\rm p}_f+\frac{1}{16\pi \ell_p^2} \frac{{\rm p} _{\phi}^2}{f {\rm p}_f}\right). 
\end{equation}
This expectation is confirmed by the analysis of the Hamiltonian flow generated by ${\rm H}_{\rm di}$ in the relevant regime. One might be slightly uncomfortable with the UV cut-off used in devising conditions such as \eqref{validity 2}, after all the classical equations predict a singular evolution when $r\to 0$ and in this limit the right hand sides of the equations \eqref{validity} actually blows up invalidating in appearance the perturbative analysis. We will see that the quantum dynamics across the classical singularity is actually well defined. Moreover, we will also see that there are exact effective equations describing the evolution of the expectation value of semiclassical states across $r=0$ as well. These equations imply that the counterpart of the right hand side of \eqref{validity} not only does not diverge but rather tends to zero when quantum effects are included. 

\subsection{Validity of perturbation theory in view of applying the model to Hawking pairs produced by a macroscopic black hole }\label{vhr}

Here we argue for the validity of the perturbation theory analysis of the previous section in the context of applications of the KS model to the
quantum dynamical description of scalar field excitations falling inside a Schwarzschild black hole during Hawking evaporation. The assumptions here are:  First the usual assumption that the gravitational collapse has formed a spherical black hole with mass $M\gg m_p$. Second, that the Hawking pairs produced by such a macroscopic black hole can be described as test field excitations on a stationary Schwarzschild geometry far away from the singularity.  Finally, we restrict our attentions to spherically symmetric Hawking pairs which are the only that can be mapped to the spherically symmetric KS configurations. This is not a serious restriction given that most of the Hawking radiation is emitted in such modes \cite{Page:1976df}.

The stationarity of the background implies that the (test field) current $j_a\equiv T_{ab}\xi^b$ is conserved, where $\xi^a$ is the stationarity Killing field that in the adapted coordinates that we use here is simply given by $\xi=\partial_t$ and $T_{ab}$ is the energy momentum tensor of the scalar field\footnote{The Hawking effect is a quantum process and the relavant state describing it is a quantum state that corresponds to the vacuum in the far past idealized by $\sI^-$. Such state can be viewed in the interior as a superposition of particles. Actual particles appear inside in the hypothetical situation of the detection of a partner at $\sI^+$ which, according to the standard interpretation of quantum mechanics, will produce a collapse of the vacuum to a new state containing an actual particle falling into the black hole. The situation after such collapse is the semiclassical situation that we model here with our classical language. A precise description of such a situation in quantum terms is a question that can only be addressed in a full theory of quantum gravity. We argue here that our little solvable model goes a humble little step into that direction.}. Conservation of energy leads to the expectation that when a Hawking particle is detected at $\sI^+$ with a given energy $\cal E=\omega$ (which coincides with the associated Killing conserved quantity), a Hawking partner with (Killing) energy $-\omega$ falls into  the black hole singularity reducing the mass of the black hole by $\omega$.
A more precise field theoretical expression of this is that the flux of the (expectation value of) the energy-momentum current $j_a$ at infinity in the state of the field after the detection of the particle at $\sI^+$ equals minus the flux of the same current on (for instance) a constant $r<2M$ hyper-surface $\Sigma_{r}$. Namely,   
\ba
\int_{\Sigma_r} j_an^a dV^{\va (3)}&=&-{\omega}.\ea
Using our coordinates to write explicitly the integrand in the left hand side while pushing $\Sigma_r$ to the region where comparison with the KS regime is possible (a constant $r$ such that $r\ll 2M$) we get
\be\label{boundy}
8\pi M   \int_0^{\ell_0} T_{r0} (t,r) r dt= 8\pi M \int_0^{\ell_0} \partial_t \phi \partial_r \phi r dt=\frac{\omega {\rm p}^2_\phi}{16\pi M\ell_0} \log(r)+\sO(1)=-{\omega},\ee
where the right hand side of the last equality comes from the evaluation of the energy flux of a Hawking particle, while in the evaluation of the left hand side
we used that for  $r\ll 2M$ (see for instance \eqref{radio})  the scalar field behaves like 
\ba \phi&=&{\rm Re}\left[e^{-i\omega t} \left(\phi_0-\frac{{\rm p}_{\phi}}{8\pi M\ell_0} \log(r)\right)\right]+\sO\left(\frac{r}{M}\right)\nn \\
&=& \cos(\omega t) \left(\phi^R_0-\frac{{\rm p}^R_{\phi}}{8\pi M\ell_0} \log(r)\right)+\sin(\omega t) \left(\phi^I_0-\frac{{\rm p}^I_{\phi}}{8\pi M\ell_0} \log(r)\right)+\sO\left(\frac{r}{M}\right).
\ea
The result in \eqref{boundy} follows also from the assumption that $\ell_0>\omega^{-1}\approx M$. By pushing the integral to its largest possible value when $r\to \ell_p$, we conclude that energy conservation (encoded in \eqref{boundy}) implies the bound \be
\frac{{\rm p}^2_\phi}{16\pi M^2} \log\left(\frac{\ell_p}{M}\right)<1,
\ee
which is consistent with the condition for the validity of perturbation theory \eqref{validity}. We see that the physical context provided by the problem of Hawking evaporation precisely justifies the simplifying assumptions that led to \eqref{Hc 2}. Notice also that the fiducial length scale $\ell_0$ acquires in such a physical situation an operational meaning as well.

\begin{figure}[h!!!!!!!]
	{\includegraphics[height=6cm]{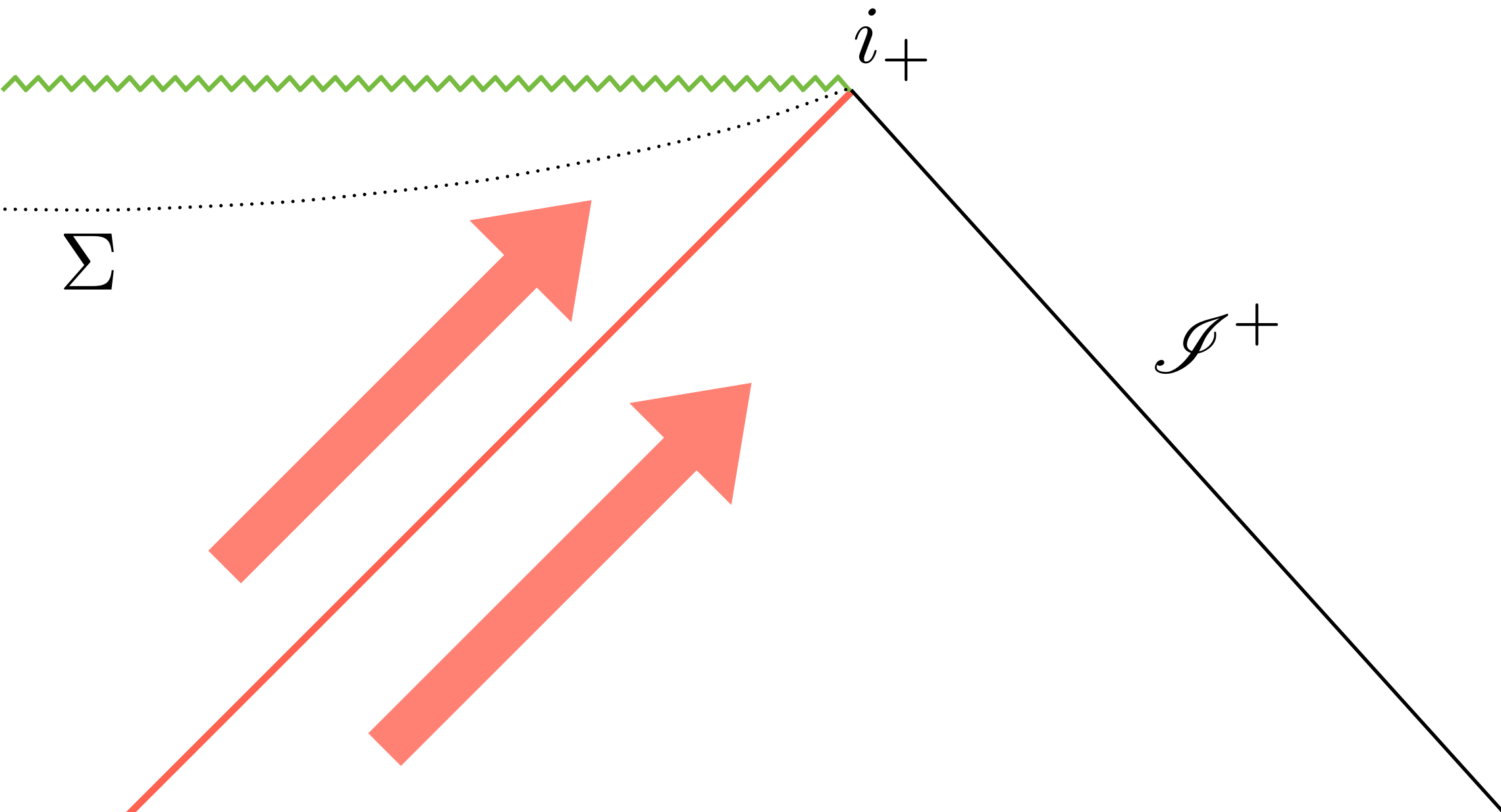}}
	\caption{Hawking pairs produced in a black hole spacetime with $M\gg m_p$ are described as test field excitations on a background spacetime that is idealized by a stationary black hole solution in the region of the Penrose diagram away from the collapsing matter and containing $i_+$. Stationarity implies the existence of a conserved energy-momentum current used, in this paper, to relate the value of ${\rm p}_{\phi}$ in the KS model to the frequency of the emitted particle at $\sI^+$ and (via Hawking temperature) to the black hole mass $M$.}.
	\label{AA-MB}
\end{figure}

A generalization of the analysis of the present Sections \ref{KSM} to a wider class of close to stationary  black holes is certainly very appealing. Even though it is clear that the tools employed here would not apply to black hole spacetimes with angular momentum (due to the breaking of spherical symmetry), one could entertain the possibility of including an electric charge without leaving the general framework of this work. However, such an apparently simple extension will reserve challenging new aspects. The first qualitative new feature is that the singularity becomes timelike in the unperturbed black hole. However, when a scalar field perturbation is added its back reaction is expected to produce  the phenomenon of mass inflation \cite{Poisson:1990eh} near the Cauchy horizon of the unperturbed background. This completely changes the global features of a realistic charged black hole interior in a way that would seem to preclude the applicability of the present methods. More precisely, if we use the Reissner-Nordstrom background black hole geometry as a basis for the present discussion---noting that the presence of mass inflation already implies that strong deviations from the Reissner-Nordstrom interior solution are to be expected---the true would-be-singularity should materialize near the location of the Cauchy horizon. However, close to the Cauchy horizon scalar field modes with frequency $\omega$ are expected to be infinitely blue shifted precluding the type of approximation available in the present case. This makes such exploration highly non-trivial even when certainly interesting.         
   
\section{A natural polymer quantization of the deep interior dynamics}

In previous sections we have shown that the Hamiltonian constraint \eqref{Hc 2} describes the deep interior dynamics of scalar excitations without angular momentum falling into a macroscopic Schwarzschild black hole. The approximations used are based on assumptions that are satisfied by the (spherically symmetric) Hawking excitations produced during black hole evaporation. Thus, the simplified dynamics in the deep interior region, which will turn out to be analytically solvable (both at the quantum and classical level), offers a toy scenario to analyze key questions of black hole evaporation in a controlled scenario.  We will show in this section that there is natural polymer quantization of the deep interior dynamics with remarkable simple properties such as: (like in the full theory of LQG) the discreteness of the area of constant area surfaces, a well defined quantum dynamics across the singularity, an effective classical description, and a direct (to our knowledge novel) link with the continuum representation. The polymer quantization we propose does not suffer from the usual ambiguities associated to the so-called holonomy corrections  \cite{Amadei:2022zwp} as the Hamiltonian constraint evolution has (in our simple model) a clear-cut geometric interpretation that allows for a unique polymerization that is compatible with the continuum limit (defined by the Schroedinger representation).  This special geometric property arises from the fact that the Hamiltonian constraint is linear in the momentum conjugate to the area of the spheres whose spectrum is discrete in our polymer representation. Ambiguities remain in the form of the so-called inverse volume corrections which are necessary if one defines the quantum dynamics across the singularity.  

\subsection{Sketch of the Schrodinger quantization} \label{q geo}

In the standard Schrodinger representation one would quantize the phase space of Section \ref{ps} by promoting the variables $a, m, {\rm p}_a, {\rm p}_m$ to self adjoint operators
\ba\!\!\!\!\!\!\!\!	\begin{array}{ccc}
	&& \widehat m\ \psi(m, {\rm p}_\phi, a) = m \psi(m, {\rm p}_\phi, a), \\ && \widehat {\rm p}_{m} \psi(m, {\rm p}_\phi, a)=-i{\partial_m \psi(m, {\rm p}_\phi, a)}, \end{array}\ 
	\begin{array}{ccc} && \widehat a\ \psi(m, {\rm p}_\phi, a)=a\psi(m, {\rm p}_\phi, a),  \\ 
	  && \widehat{{\rm p}}_a \psi(m, {\rm p}_\phi, a)=-i {\partial_a \psi(m, {\rm p}_\phi, a)},\end{array} \  
	\begin{array}{ccc} && \widehat {\rm p}_\phi \psi(m, {\rm p}_\phi, a)={\rm p}_\phi \psi(m, {\rm p}_\phi, a),  \\ 
	  && \widehat{\phi}\ \psi(m, {\rm p}_\phi, a)=i {\partial_\phi \psi(m, {\rm p}_\phi, a)},\end{array} \nn
\ea
in the kinematical Hilbert space is $\sH_{\rm S}={\sL}^2(\mathbb{R}^3) $, equipped with the usual inner product
\begin{equation} \label{usual ip}
	\langle \psi_1 , \psi_2 \rangle = \int_{-\infty}^{+\infty}\int_{-\infty}^{+\infty} \int_{-\infty}^{+\infty}   \overline{\psi_1(m, {\rm p}_\phi, a)} \psi_2(m, {\rm p}_\phi, a)  dm d{\rm p}_\phi  da  ,
\end{equation}
where we have chosen the momentum representation for the scalar field for convenience (as ${\rm p}_{\phi}$ is one of the constants of motion of the system).
Eigenstates of the $\widehat a$ operator are interpreted as distributions (they are not in the Hilbert space) and one usually writes
\be
\widehat a \ket{a}=a \ket{a}
\ee
with $a\in \R$ and form an orthonormal basis 
\be\label{kikin}
\braket{a,a'}=\delta({a,a'}).
\ee
The dynamics is imposed by solving the Hamiltonian constraint (\ref{Hc 2}) which, in the present representation, takes the precise form of a Schrodinger equation in the area variable $a$, namely 
\be\label{SE}
\left[-i\hbar \frac{\partial}{\partial a}+\frac{1}{2a}\left(m+\frac{{\rm p}_{\phi}^2}{16\pi \ell_p^2 m}\right)\right]\psi(m, {\rm p}_\phi, a)=0.
\ee
As usual, solutions of the constraint are certainly not square integrable in the $a$-direction, thus physical states are outside of the kinematical Hilbert. The physical Hilbert space is defined as the space of square integrable functions of $m$ and ${\rm p}_{\phi}$ at fixed time $a$---$\sH_{\rm phys}={\sL}^2(\mathbb{R}^2)$---with inner product
\be
\label{usual phys ip}
	\langle \psi_1(a), \psi_2(a) \rangle_{\rm phys} = \int_{-\infty}^{+\infty} \int_{-\infty}^{+\infty}   \overline{\psi_1(m, {\rm p}_\phi, a)} \psi_2(m, {\rm p}_\phi, a)  dm d{\rm p}_\phi ,
\ee 
which is preserved, i.e. it is independent of $a$,  by the Schrodinger equation (evolution is unitary in $a$).

Two important remarks are in order: 
First note that we are formulating in detail the dynamics of the system in the near singularity approximation. The physical reason for this is that (as argued previously) it is only in this approximation that the system can be compared with a (spherically symmetric) black hole with spherically symmetric excitations falling inside. A side gain is also the simplification of the dynamics which will allow us for a simpler quantization and the analysis of the possibility of a well defined dynamics across the singularity when we undergo the LQG inspired quantization. One could however consider the quantization of the minisuperspace system without the near singularity approximation. In that case one would need to write a Schroedinger equation using the Hamiltonian \eqref{wewe}, now genuinely time-dependent ($r$-dependent), for which unitary evolution would involve path ordered exponentials (as the Hamiltonian does not commute with itself at different $r$ values). In addition one would need to work with either $r,f, {\rm p}_r, {\rm p}_f$ variables or $a,f, {\rm p}_a, {\rm p}_f$ variables without the luxury of the simplifications introduced by the use of the near singularity variables \eqref{nsv}.    

 \subsection{The polymer quantization}
 
We define now a representation of the phase space variables that incorporates a key feature of the full theory 
of LQG: the area quantization. Such representation closely mimics the structure of the quantum theory in the fundamental theory in such a way that the area variable $a$ acquires a discrete spectrum. Mathematically, this is achieved by replacing the $\sL^2$ structure of the inner product in the variable ${\rm p}_a$ by the inner product of the Bohr compactification of the ${\rm p}_a$ phase space dimension. More precisely one  substitutes the kinematical inner product in the Schroedinger representation \eqref{usual ip} by
\begin{equation} \label{bohr ip}
\langle \psi_1 , \psi_2 \rangle  = \lim_{\Delta \rightarrow +\infty} \frac{1}{2\Delta}\int_{-\Delta}^{+\Delta}  \left(\int_{-\infty}^{+\infty} \overline{\psi_1(m, {\rm p}_\phi, {\rm p}_a)} \psi_2(m, {\rm p}_\phi, {\rm p}_a)  dm d{\rm p}_\phi  \right)d{\rm p}_a \ .
\end{equation}
With this inner product, periodic functions of ${\rm p}_a$ with arbitrary period are normalizable and the conjugate $a$-representation acquires the property of discreteness in a way that closely mimics the structure of the fundamental  theory of loop quantum gravity \cite{Ashtekar:2011ni}. In particular eigenstates of $
\widehat a$ exist 
\be
\widehat a \ket{a}=a \ket{a}
\ee
with $a\in \R$. These state form an orthonormal basis with inner product 
\be
\braket{a,a'}=\delta_{a,a'},
\ee
in contrast with \eqref{kikin}.
Discreteness of the spectrum of $\widehat a$ comes at the prize of changing the kinematical Hilbert space structure in a way that precludes the infinitesimal translation operator $\widehat {\rm p}_a$ to exist. Instead only finite translations (quasi periodic functions of ${\rm p}_a$) can be represented as unitary operators in the polymer Hilbert space. Their action on the $a$-basis is given by
\begin{equation}\label{defidefi}
	\widehat{e^{i\lambda {\rm p}_a}}\psi(m, {\rm p}_\phi, a)=\psi(m, {\rm p}_\phi,a+\lambda \ell_p).
\end{equation}
Eigenstates of the finite translations (or shift operators)  exist and are given by wave functions supported on discrete $a$-lattices. Namely,
\be \label{cocuna} \psi_{k,\epsilon}(a)\equiv \left\{ \begin{array}{ccc} &\exp (i k a)& \ \ \ {\rm if} \ \ \ a\in \Gamma_{\epsilon,\lambda}\equiv \{(\epsilon+n\lambda)\ell_p^2\in \R \}_{n\in \mathbb{Z}} \\ &0& \ \ \ {\rm otherwise}
\end{array}\right.\ee
where  the parameter $\epsilon\in [0,\lambda)\in \R$. The discrete lattices denoted $\Gamma_{\epsilon,\lambda}$ are the analog of the spin-network graphs in LQG with the values of $a$ on lattice sites the analog of the corresponding spin labels.  With all this one has  (using \eqref{defidefi}) that
\be
\widehat{e^{i\lambda {\rm p}_a}}\psi_{k,\epsilon}(a)= {e^{i\lambda k}} \psi_{k,\epsilon}(a).
\ee
Note that, unlike the Schroedinger representation where the eigen-space of the momentum operator is one dimensional, the eigen-spaces of the translation operator (labelled by the eigenvalue ${e^{i\lambda k}}$) are infinite dimensional and non separable. This is explicit from the independence of the eigen-values of the continuous parameter $\epsilon\in [0, \lambda)$ labelling eigenstates. Such huge added degeneracy in the spectrum of the shift operators is a general feature of the polymer representation. We will show that this degeneracy can show up in Dirac observables of central physical importance such as the mass operator in Section \ref{degeM}. 

\subsection{Quantum dynamics}
 The dynamics is dictated by the quantization and imposition of the constraint \eqref{Hc 2}. As the operator corresponding to ${\rm p}_a$ does not exist in our kinematical Hilbert space we introduce a polymerized version. Traditionally, this is achieved by replacing the infinitesimal translation operator \be \lambda \widehat {\rm p}_a \ \ \ \longrightarrow_{\!\!\!\!\!\!\!\!\!\!\!\!\!\!\!\!{}_{\rm traditional}} \ \ \ {\widehat{\sin{\lambda {\rm p}_a}}}.\ee
 The rule consists of making some `minimal' substitution of ${\rm p}_a$ by a periodic regularization satisfying that in the limit $\lambda\to 0$ the functional choice will approximate the original function.
Such rule is intrinsically ambiguous and,  it opens in general the door for an infinite set of possibilities. Such choices are to be interpreted as quantization ambiguities of the Hamiltonian constraint with potential quantitative dynamical consequences (for a general discussion see \cite{Perez:2005fn}). Dynamical implications of these ambiguities can be analysed in detail in simple models of quantum cosmology \cite{Amadei:2022zwp} and black holes \cite{Munch:2022teq}.   

In the full theory a new perspective on the regularization issue has been introduced 
motivated by the novel mathematical notion of generalized gauge covariant Lie derivatives \cite{Ashtekar:2020xll} and their geometric interpretation allowing for the introduction of a natural regularization (and subsequent) anomaly free quantization of the Hamiltonian constraint \cite{Varadarajan:2022dgg}. Even when the procedure  does not eliminate all ambiguities of quantization (choices are available in the part of the quantum constraint responsible for propagation \cite{ale-madhavan}), the new technique reduces drastically some of them in the part of the Hamiltonian that is more stringently constrained by the quantum algebra of surface deformations.  

What we want to emphasize here is that the analogous procedure in the case of our symmetry reduced Hamiltoinian has a similar effect. Indeed, because our classical Hamiltonian constraint is linear in the variable ${\rm p}_a$ (whose associated Hamiltonian vector field has a crystal-clear geometric interpretation of infinitesimal translations in $a$), one has an unambiguous choice of quantization: the obvious choice to replace infinitesimal translations (which do not exist in the polymer representation) by finite translations or shifts. From this geometric perspective the right polymerization is the regularization that make the replacement
\be \lambda \widehat {\rm p}_a \ \ \ \longrightarrow \ \ \ \widehat{e^{i\lambda {\rm p}_a}}.\ee
In other words the differential time evolution in the Schrodinger equation must be represented in the polymer Hilbert space by a finite translation with a polymerization scale $\lambda$. However, as such an action is associated with a clear geometric meaning, the geometric compatibility with the Schrodinger equation can be preserved if the second term in the classical Hamiltonian \eqref{Hc 2} is exponentiated too in order to produce the well known unitary evolution operator that produces finite area evolution. Disregarding for the moment quantum corrections that will have to be included near the $a=0$ region (see Section \ref{qes}), the quantum constraint is taken to be
\be\label{dydy}
{\underbrace{\exp({i\lambda {\rm p}_a})}_{\begin{array}{ccc} \rm \van finite \ areatime \\ \rm \van translation\end{array}}\ket{\psi} -  \overbrace{\exp\left({\frac{i}{2} \log\left(\frac{a+\lambda \ell^2_p}{ a}\right) \left(m+\frac{{\rm p}_{\phi}^2}{16\pi \ell_p^2 m}\right)}\right) }^{\begin{array}{ccc} \rm \van finite \ areatime \\ \rm \van unitary \ evolution \ operator\end{array}}}\ \ket{\psi} =0, 
\ee 
whose action is well defined in the polymer representation and whose solutions are easily found (by acting on the left with $\bra{m,{\rm p}_\phi,a}$) to be wave functions satisfying the discrete dynamics given by
\be\label{step}
\psi(m,{\rm p}_\phi, a+\lambda \ell_p^2)
=e^{\frac{i}{2} \log\left(\frac{a+\lambda \ell^2_p}{ a}\right) \left(m+\frac{{\rm p}_{\phi}^2}{16\pi \ell_p^2 m}\right)}  
\psi(m,{\rm p}_\phi, a).
\ee
The physical Hilbert space is defined via the usual inner product at fixed (discrete) time $a$ via
\be
\label{phys ip}
	\langle \psi_1(a), \psi_2(a) \rangle_{\rm phys} = \int_{-\infty}^{+\infty} \int_{-\infty}^{+\infty}   \overline{\psi_1(m, {\rm p}_\phi, a)} \psi_2(m, {\rm p}_\phi, a)  dm d{\rm p}_\phi ,
\ee   
which independent of the lattice sites  as required (a property that we could identify with the intrinsic unitarity of the quantum constraint kernel). More precisely the physical inner product is a constant of the quantum motion associated to the full history represented by the lattice $\Gamma_{\epsilon, \lambda}$ as implied by
unitarity.  Explicitly one has
\be
\langle \psi_1(a), \psi_2(a) \rangle_{\rm phys}= \langle \psi_1(a+\lambda), \psi_2(a+ \lambda) \rangle_{\rm phys}\ \ \ \forall \ \ \ a\in \Gamma_{\epsilon, \lambda}.
\ee
Ambiguities of regularization that are usually associated with the polymerization procedure are thus completely absent in this model. The reason is the linear dependence of the Hamiltonian constraint in the polymerized variable which allows for a regularization fixed by the geometric interpretation of the classical Hamiltonian vector field associated to the corresponding variable. However, ambiguities remain  when one studies the evolution across the {\em would-be-singularity} of the Kantowski-Sachs model at $a=0$. We will study this in the next section.   

Now we would like to concentrate on the evolution when we are away from the $a=0$.
In such regime the one step evolution \eqref{step} can be composed to produce the arbitrary initial to final area evolution 
\be
{\psi(m,{\rm p}_\phi, \epsilon+n \lambda )=
 \left( \frac{\epsilon+n \lambda}{ \epsilon+q \lambda}\right)^{\frac{im}{2} \left(1+\frac{{\rm p}_{\phi}^2}{16\pi \ell_p^2 m^2}\right)}  
\psi(m,{\rm p}_\phi, \epsilon+q \lambda)},
\ee
for arbitrary integers $n,q>1$. Evolving across the $a=0$ point will be discussed later.

\subsection{The continuum versus the polymer dynamics}

The polymer dynamics that arises from the geometric action of the quantum constraint \eqref{dydy} enjoys of the appealing feature
of being closely related to the dynamics that one would obtain in the continuum Schroedinger representation. This statement can be made precise as follows:  any solution of the Schroedinger equation
\eqref{SE} induces on any given lattice $\Gamma_{\epsilon, \lambda}$ a solution of \eqref{dydy}. Conversely, physical states of the polymer theory represent a discrete sampling of the continuum solutions of \eqref{SE}. However, the Schroedinger evolution is ill defined at the singularity $a=0$ due to the divergence of the $1/a$ factor in front of the second term of \eqref{SE}. The polymer representation allows for a well defined evolution across the singularity thanks to the deviations from the $1/a$ behaviour introduced by the analog of the `inverse-volume' corrections (see next section). 
Nevertheless, with the appropriate modification of the $1/a$ factor in the Schroedinger equation the  
correspondence between the discrete and continuum continues to hold.

\subsection{Quantum evolution across the classical {\em singularity}}\label{qes}

\begin{figure}[h!]
	{\includegraphics[height=8cm]{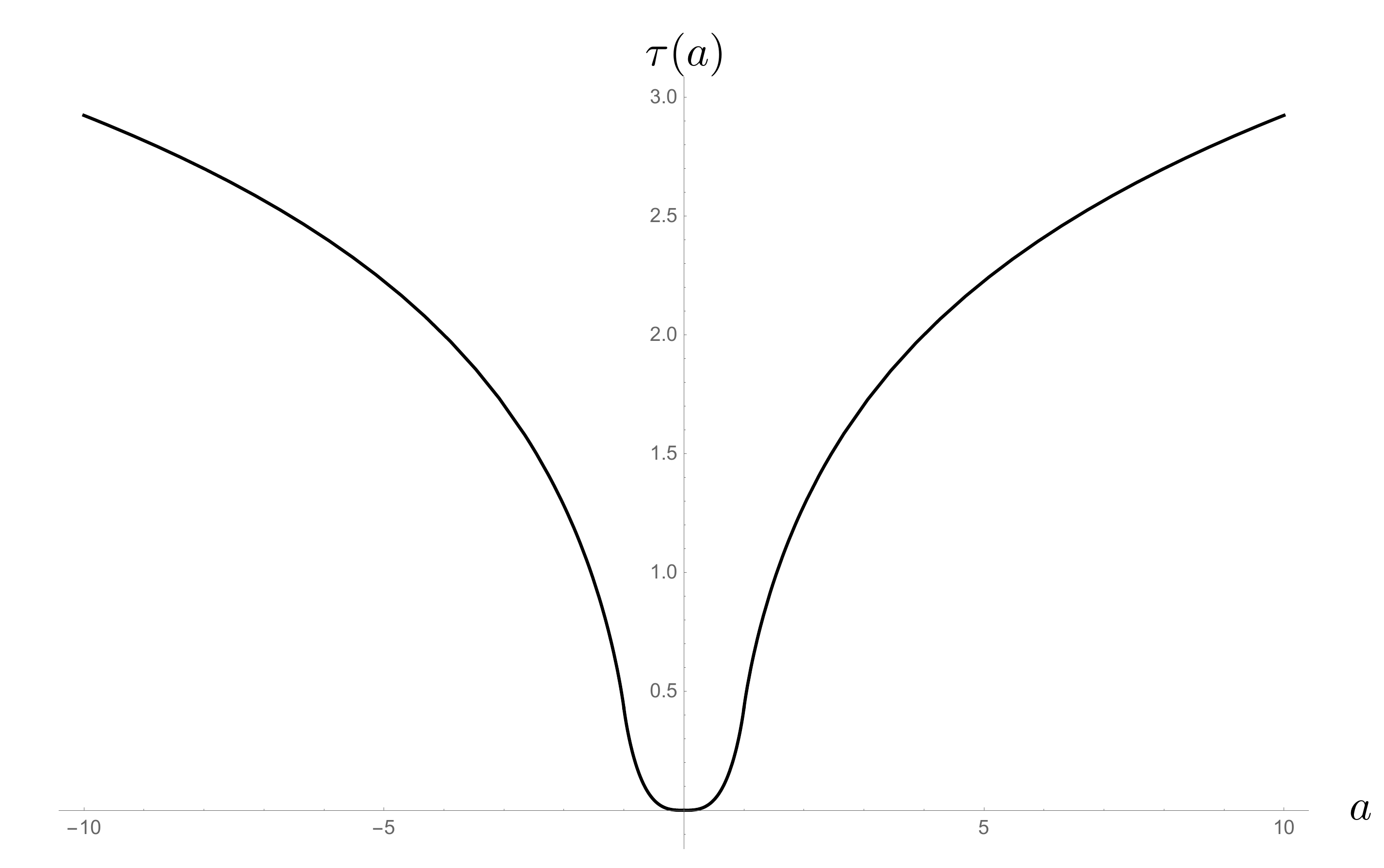}}
	\caption{The dynamics across the singularity is regular due to inverse volume corrections. In this picture we show the function $\tau(a)$, defining the dynamical evolution across the singularity, in the case of Thiemann's regularization of the inverse-area-operator. }.
	\label{grey}
\end{figure}

Quantum evolution in $a$ for all values of $a$ including the singularity is dictated by the quantum corrected version of the constraint \eqref{dydy} given by 
\begin{align}\label{step1}
	\psi(m, {\rm p}_{\phi}, a) &=  e^{\frac{i}{2} \left[ \int\limits_{a_0}^{a}\left[\frac{1}{a}\right]_qda \right]\left( m + \frac{{\rm p}_{\phi}^2}{16\pi \ell_p^2 m} \right) } \psi(m, {\rm p}_{\phi}, a_0) \\
	&= e^{\frac{i}{2} \left[\tau(a)-\tau(a_0)\right] \left( m + \frac{{\rm p}_{\phi}^2}{16\pi \ell_p^2 m} \right) } \psi(m, {\rm p}_{\phi}, a_0),
\end{align}
where 
\be
\left[\frac{1}{a}\right]_q
\ee
denotes the quantum corrected expression for the operator $a^{-1}$ (the analog of inverse volume correction in cosmology) that can be implemented in various ways due to inherent ambiguities associated to the polymer quantization. In general this will give deviations of the $a^{-1}$ behaviour in the region $a\sim \ell_p^2$. This modifies the integral of $a^{-1}$ in a way charaterized by the (to a large extend arbitrary \cite{Amadei:2022zwp}) function $\tau(a)$ introduced in the second line.
One, among the many possibilities, is the one that follows from the so-called Thiemann's trick whose most elementary form is (see \cite{Ashtekar:2011ni} for its application in cosmology, see \cite{Amadei:2022zwp} for a discussion of the multiplicity of variants)
\be\label{regulo}
\left[\frac{1}{a}\right]_{\rm Thm}\equiv{\rm sign}(a) \left(\frac{\sqrt{|a+\ell_p^2|}-\sqrt{|a-\ell_p^2|}}{\ell^2_p}\right)^2=\frac{1}{a}+\sO\left(a^{-3}\right).\ee  
Integration leads to the following $\tau(a)$ function in \eqref{step1}  
\be
\tau(a)=\left\{\begin{array}{ccc} |a|
   \left(|a|- \sqrt{|a|^2-1}\right)+\log
   \left(\sqrt{|a|^2-1}+|a|\right)
   -\frac{\pi }{2}+1&\ \  \  1\le |a| \\
-|a|
   \left(\sqrt{1-|a|^2}-2\right)
   -\sin ^{-1}(|a|)& \ \  \  |a|\le 1 \end{array}\right. ,
\ee
whose graph is presented in Figure \ref{grey}.


The classical theory does not fix the evolution uniquely in this high curvature regime where quantum geometry effects cannot be neglected. Quantum geometry effects regularize the dynamics near the $a=0$ singularity; one way of seeing this is that the factor $\log(a)$ in the quantum evolution away from the singularity in \eqref{dydy} receives inverse volume quantum geometry corrections. As  discussed in \cite{Amadei:2022zwp} these corrections are ambiguous (as fact that should not be surprising given the expectation that the classical theory cannot guide us all the way to the deep UV in QFT). Instead of proposing one particular UV extension, as in the example shown where Thiemann regularization was used, one might simply keep all possibilities open and assume that the corresponding operator  is regularized in the relevant region by some arbitrary function $\log(a)\to \tau(a)$. In regions where $\tau(a)=\log(a)$ the quantum evolution leads to semiclassical equations that match exactly Einstein's equations in the KS sector (more details in Section \ref{semisemi}) 

 \subsection{Dynamics of semiclassical states and tunelling across the singularity}\label{semisemi}

Let us first consider the vacuum ${\rm p}_{\phi}=0$ case. This case is important because it should correspond to the 
Schrwarzschild interior in the region $r\ll M$ (or $a\ll M^2$). The dynamical evolution \eqref{step1} becomes
\begin{align}
	\psi(m, a) &= e^{\frac{i}{2}m (\tau(a)-\tau(a_0))} \psi(m, a_0).
\end{align}
In the $p_{\rm m}$ representation the previous equation simply reads
\begin{align}\label{sisi}
	\psi(p_{\rm m}, a) &= \psi(p_{\rm m} + \frac{\tau(a)-\tau(a_0)}{2}, a_0),
\end{align}
i.e., a simple translation in momentum space.
Such  dynamical evolution in $a$ implies in an obvious manner that semiclassical states peaked at classical values $(\overline m, \overline {\rm p}_{m})$ at area $a_0$
with some given fluctuations will be simply evolved into the translated state with the very same fluctuation properties 
peaked at $(\overline m, \overline {\rm p}_{m} + (\tau(a)-\tau(a_0))/2)$ at area $a$. 
This is a remarkable property of the quantum system:  independently of the quantum gravity effects encoded in the precise form of $\tau(a)$,  expectation values satisfy well defined effective dynamical equations which are exact (not an approximation), and semiclassical states are not spread by the dynamical evolution (as in the simple case of the harmonic oscillator). Notice that the validity of effective dynamical equations in the context of black hole models have remained a conjecture in other formulations \cite{Ashtekar:2018cay}.

In regions where $\tau(a)=\log{a}$ the previous evolution gives
\be
\overline {\rm p}_{m}(a)=\log(r)+\overline {\rm p}_{m}(a_0), \ \ \ \overline m(a)=\overline m(a_0).
\ee
From the definitions \eqref{nsv} we have that $f p_f= m(a)$ and $f=-\exp[- {\rm p}_{m}(a)]$.  The constraint \eqref{central}, gives us  $h(r)$ and the metric becomes
	\be \label{ds(m,pm)}
	ds^2=e^{- {\rm p}_{m}(a)} 
	dt^2 -\frac{\ell^2_0}{(4\ell_p)^4\pi^2}
	\frac{e^{- {\rm p}_{m}(a)}}{ m(a)^2} da^2 
	+\frac{a}{4\pi} d\Omega^2
	\ee
	which corresponds to the Schrwarzschild solution with the two Dirac observables $M$ and ${\rm p}_{M}$ given in terms of the initial conditions by  \be M= \frac{2\sqrt{4\pi}\ell_p^4}{\ell^2_0\sqrt{a_0}}  m(a_0)^2 e^{{\rm p}_{m}(a_0)} ,\ \ \  \ \ \  {\rm p}_{M}=  \frac{\ell_0\sqrt{4\pi}}{2\ell_p \sqrt{a_0}}
\frac{e^{- \overline {\rm p}_{m}(a_0)}}{\overline m(a_0)}.\ee  
When quantum inverse volume corrections are taken into account then the quantum evolution is perfectly well defined across the classical singularity. The evolution of the mean values of a semiclassical state is also well defined and given by
\be
\overline {\rm p}_{m}(a)=\overline {\rm p}_{m}(a_0)+\frac{1}{2}(\tau(a)-\tau(a_0)), \ \ \ \overline m(a)=\overline m(a_0). 
\ee
Such solutions can also be obtained from the Hamiltonian constraint given that one replaces the operator $a^{-1}$ by its quantum regularization (effective equations are in this sense exact). The metric for all values of $|a|\ll M^2$ but otherwise arbitrary becomes
\be\label{memetri}
ds^2={\exp\left[{-\overline {\rm p}_{m}(a_0)-\frac12 ({\tau(a)-\tau(a_0)})}\right]} 
\left( dt^2 -\frac{\ell^2_0}{(4\ell_p)^4 \pi^2}
\frac{da^2}{\overline m(a_0)^2}  \right)
+\frac{a}{4\pi} d\Omega^2.
\ee
Since the Thiemann regularization produces a $\tau(a)\sim a^2\sim r^4$ near $a=0$ we see that the previous metric is just given by a two dimensional Minkowski metric fibrated with two dimensional sphere with time dependent radius $r$. The shrinking of the spheres leads to a singularity at $a=0$ where the spheres collapse and the spacetime geometry (as described by the effective line element) becomes a two dimensional flat one at the singularity in the $a$-$t$ plane. Despite the singular nature of the effective metric the fundamental quantum evolution is well defined across the singularity.    

In the presence of matter the situation is a bit more involved due to the factor ${\rm p}_{\phi}/ m$ 
appearing in matter contribution to the Hamiltonian constraint \eqref{dydy}.
However, in the spirit of applying this analysis to macroscopic black holes and modelling the dynamics of a 
weak scalar excitation (a Hawking particle) falling into the singularity, it is natural to focus on semiclassical states (Gaussian)
peaked on values such that $\overline m\gg {\rm p}_{\phi}$ with fluctuations $\sigma_m\ll \overline m$.
As in the vacuum case $\overline m=\overline m_0$, i.e., $\overline m$ is a constant of motion as well as its spread $\sigma_m$. The dynamics of the conjugate variable
(the mean value $\overline {\rm p}_m$ of the variable ${\rm p}_m$) can be evaluated using stationary phase methods and the result is 
\be\label{pipipa}
\overline {\rm p}_{m}(a)=\overline {\rm p}_{m}(a_0)+\frac{1}{2}(\tau(a)-\tau(a_0))\left[ 1+ \frac{{\rm p}_{\phi}^2}{\overline m(a_0)^2} \left(1 +\frac 34\frac{\sigma_m^2}{\overline m(a_0)^2}\right)\right]+\sO\left(\frac{\sigma_m^3}{\overline m(a_0)^3}\right),
\ee
which can be seen to correspond to the classical solutions found in Section \ref{solubili}. Notice that, as expected from the form of the matter coupling, there are here quantum corrections characterized by terms proportional to ${\sigma_m^2}/{\overline m(a_0)^2}$.
The spread in the variable ${\rm p}_{m}$ is not time independent if we take into account higher order corrections, namely
\be
\overline \sigma^2_{{\rm p}_{m}}(a)=\frac{1}{\sigma^2_{m}} + \frac{\sigma^2_{m} {\rm p}_{\phi}^4}{4 \overline m(a_0)^4} (\tau(a)-\tau(a_0))^2.
\ee
The previous equations are derived assuming that the scalar field is in an eigenstate of the momentum ${\rm p}_{\phi}$. This is an idealization that simplifies the analysis of the dynamical evolution of the geometry. Similarly, if we assume that the geometry state was in an eigenstate of $m$ then we can easily analyze the dynamics of the scalar field assuming that it is initially in a gaussian semiclassical state picked about $\overline {\rm p}_{\phi}(a_0)$ and $\overline \phi(a_0)$. In accordance with the classical solutions we get
\ba
&& \overline {\rm p}_{\phi}(a)=\overline {\rm p}_{\phi}(a_0) \nn \\
&& \overline{\phi}(a)=\overline{\phi}(a_0)+\left(\tau(a)-\tau(a_0) \right) \frac{\overline {\rm p}_{\phi}(a_0)}{16\pi \ell_p^2 m}.
\ea
One way to quickly  derive these equations by inspection is to realize that the Hamiltonian constraint \eqref{Hc 2} is that of a non relativistic point particle 
with mass proportional to our geometric variable $m$ evolving in $d\tau=[1/a]_q da$. Note that \eqref{pipipa} implies that the back-reaction of the scalar field enters only through a simple modification of the exponential conformal factor in front of the $2$-metric in the $a$-$t$ `plane' in equation \eqref{memetri}.

Unlike the geometry degrees of freedom, the fluctuations in the scalar field grow as one approaches the would-be-singularity: for a given geometry semiclassical state picked around the mass $M$, the spread of the scalar field $\sigma_\phi$ in an initially eigenstate of $\phi$ at area $a$  grows to a maximum value close to the would-be-singularity such that \be M \sigma_\phi< \sqrt{\frac{M \log\left({a}/{\ell_p^2}\right)}{16\pi \ell_0}}, \ \ \ \sigma_{{\rm p}_{\phi}}< \sqrt{\frac{16\pi M }{ \ell_0 \log\left({a}/{\ell_p^2}\right)}}\ee which are small in the interior if we take $\ell_0\gg M$ as expected from appearance of $\ell_0$ in the fundamental commutation relations \cite{Rovelli:2013zaa}. 
Note that, if we take into account inverse volume corrections of the type suggested by LQG (see Figure \ref{grey}),
the scalar field reaches a critical point at $a=0$ where its area velocity vanishes.

\subsection{The mass operator (in the vacuum case)}\label{degeM}

In this section we concentrate in the vacuum case for simplicity as the mass can be directly read off the form of the metric in this case via a simple comparison with the classical Schwarzschild solution.  
In this case the result is
\begin{equation}\label{M(m,pm)}
M=\frac{2\sqrt{4\pi}\ell_p^4}{\ell_0^2\sqrt{a}}m^2e^{{\rm p}_m}.
\end{equation}
where we used the vacuum solution \eqref{schww} (in its $r\ll M$ approximation) and \eqref{nsv}.
 It is easy to verify that the previous is indeed a Dirac observable by showing that it commutes with Hamiltonian constraint  \eqref{Hc 2}. 
Its non linear dependence on the basic variables anticipates factor ordering ambiguities when it comes to promoting the mass to a quantum operator. 
Here we focus on the choice
\be\label{massa} \widehat M =\alpha(\widehat a)\left[ \widehat m {e^{\widehat{\rm p}_m}}\widehat m\right] , 
\ee
where $\alpha(a)\equiv{2\sqrt{4\pi}\ell_p^4}/{(\ell_0^2\sqrt{a})}$. The eigenstates equation
\begin{equation}
\widehat M |\phi_M\rangle=M|\phi_M\rangle,
\end{equation}
turns into the differential equation 
\begin{equation}
\alpha(a) e^{{\rm p}_m}\frac{\partial \phi_M({\rm p}_m,a)}{\partial {\rm p}_m}+ \alpha(a) e^{{\rm p}_m}\frac{\partial^2 \phi_M({\rm p}_m,a)}{\partial {\rm p}_m^2}+M \phi_M({\rm p}_m,a)=0 ,
\end{equation}
if we expand the eigenstate in the ${\rm p}_m,a$ basis, namely \begin{equation} \label{eigenstate M}
|M \rangle=\sum_{a \in \Gamma_{\epsilon,\lambda}} \int \phi_M({\rm p}_m,a)|{\rm p}_m\rangle| a\rangle d{\rm p}_m,
\end{equation}
where the sum runs over the discrete lattice $\Gamma_{\epsilon,\lambda}$ defined in \eqref{cocuna} when introducing the dynamical constraint \eqref{dydy}.
This differential eigenvalue equation is solved by
\begin{equation}\label{eigen M}
\phi_M({\rm p}_m,a)\equiv \bra{{\rm p}_m,a}M \rangle=  \sqrt{\frac{2\sqrt{M}}{\alpha(a)}}e^{-{\rm p}_m/2} J_1\left(2\sqrt{\frac{M}{\alpha(a)}}e^{-{\rm p}_m/2}\right),
\end{equation}
where $J_1$ is a Bessel function. 
One can explicitly verify that the quantum dynamics \eqref{dydy} preserves the eigenstates by explicitly showing that the evolution between arbitrary lattice points $a_1, a_2 \in \Gamma_{\lambda, \epsilon}$  sends the wave function of the eigenstate at the $a_1$ lattice point to the $a_2$ lattice point (as expected for a Dirac observable); or equivalently, the eigenstates of the mass are physical states solving \eqref{dydy}. Explicitly,   
\begin{equation}
\widehat{e^{\frac{i}{2}(\log(a_2)-\log(a_1))m}}\phi_M({\rm p}_m,a_1)=\phi_M\left({\rm p}_m+\frac{1}{2}(\log(a_2)-\log(a_1)),a_0\right)=\phi_M\left({\rm p},a_2\right).
\end{equation}
Now the evolution across $a=0$ requires inverse volume corrections which modifies the previous dynamical law by replacing $\log(a)\to \tau(a)$. The mass Dirac 
observable still exists once inverse volume corrections are turned on. It corresponds to the modification of \eqref{massa} via the substitution 
$\widehat a\to \exp({{\widehat \tau(a)}})$. Eigenstates are also obtained by the same substitution in \eqref{eigen M} and satisfy the expected 
Dirac observable condition (which now holds for lattice points at different sides across the singularity)
\begin{equation}
\widehat{e^{\frac{i}{2}(\tau(a_2)-\tau(a_1))m}}\phi_M({\rm p}_m,a_1)=\phi_M\left({\rm p}_m+\frac{1}{2}(\tau(a_2)-\tau(a_1)),a_0\right)=\phi_M\left({\rm p},a_2\right).
\end{equation}
When supported on the same lattice, one can show that they satisfy the orthogonality relation
\be
\braket{M|M^\prime}_{\rm phys}=\delta(M,M^\prime),
\ee  
where the inner product is computed with the physical inner product \eqref{phys ip}.
Thus the spectrum of the mass operator is continuous. It was argued in the context of the full LQG theory in \cite{Perez:2014xca,  Amadei:2019wjp, Perez:2022jlm} that the eigenspaces of the mass should be infinite degenerate due to the underlying discrete structure of the fundamental theory and the existence of defects that would not be registered in the ADM mass operator. Interestingly,  the conjectured property is illustrated explicitly in our simple toy model as the eigenvectors \eqref{eigenstate M} for a given eigenvalue $M$ there are infinitely many and labelled by a continuum parameter. More precisely they are associated with wave functions of the form \eqref{eigen M} supported on lattices with different values of $\epsilon$. Thus eigenstates of the mass should then be denoted $|M, \epsilon \rangle$ with orthogonality relation 
\be
\braket{M,\epsilon |M^\prime, \epsilon^\prime}_{\rm phys}=\delta(M,M^\prime)\delta_{\epsilon,\epsilon^\prime},
\ee  
where $\delta_{\epsilon,\epsilon^\prime}$ is the Kronecker delta symbol.
The existence of such a large degeneracy is a generic feature of the polymer representation. Even when this is a toy model of quantum gravity, this feature 
is likely to reflect a basic property of the representation of the algebra of observables in the full LQG context. Here we are showing that the mass operator is hugely degenerate suggesting that the usual assumption of the uniqueness of the vacuum in background dependent treatments of quantum field theory might fail in a full loop quantum gravity context. 

Alternative factor orderings of the quantum operator $ M$ 
could be treated similarly (some simple choices lead to slightly different eigenvectors written also in 
terms of Bessel functions). Such an ambiguity is not relevant for our purposes (and it does not change the key fact that the spectrum of $M$ is infinitely degenerate) as the aim of the model is not to construct any quantitative physical prediction but rather to use it as a toy model to investigate possibly sufficiently generic features that could actually survive in the full theory.
The large degeneracy of the mass spectrum is, in our view, an interesting example of one such feature. 


\section{Discussion}

We have shown that test field solutions of the Klein-Gordon equation with zero angular momentum behave like solutions of the KS symmetry reduced model in the {deep interior region} $r\ll M$ defined in terms of the background Schwarzschild spacetime. This implies that,  spherically symmetric scalar matter falling into  a spherical black hole can be modelled by the KS solutions near the singularity. Despite the expected limitations of symmetry reduced models in capturing the full physics in the UV regime, the model includes 
back-reaction of the scalar matter. Focusing in the {deep interior region} and using perturbation theory in ${\rm p}_\phi/M$ we show that it is possible to interpret  the solutions of KS with matter as Schwarzschild solutions with matter excitations falling towards the $r=0$ singularity (this interpretation is not global but it shown to be correct in the {deep interior region}). 
The Hamiltonian dynamics simplifies considerably in that regime becoming tractable both at the classical as well as the quantum level.  Perturbation theory applies (we have shown) to the situation involving Hawking particles falling into the singularity.

In close analogy to LQG, we define a quantization of the system describing the {deep interior region} where the area of the $r=$constant spheres has a discrete spectrum. This leads to the polymer representation of the area of the orbits of the rotation group and its conjugate momentum that allows for a well defined quantum evolution across the singularity if one introduces customary `inverse-volume' corrections to the quantum scalar constraint. The Hamiltonian constraint admits a simple geometric interpretation in the to-be-polymerized sector due to the linearity of the Hamiltonian constraint in the momentum variable conjugated to the area of the $r=$constant spheres. The geometric nature of the action of the classical constraint allows for the introduction of a unique polymerization prescription respecting this action at the quantum level. This reduces the ambiguity usually associated to the procedure of quantization of the dynamical constraints for reasons that resonate with the ones that lead to similar advantages in the full theory \cite{Ashtekar:2020xll}. Remarkably, the dynamics is exactly solvable at the quantum level. In the vacuum case, the mass operator is a Dirac observable that we quantize and whose spectrum is given explicitly. Semiclassical states dynamics leads to effective evolution equations that can be characterized exactly in the vacuum case and using suitable stationary phase approximations in the case where matter is present. These effective equations coincide with Einsteins equations in regions where the inverse volume corrections can be neglected. 

An important formal aspect of the model is that it presents a concrete example of violation of the `unicity of the vacuum' assumption that permeates discussions of Hawking's information puzzle for over 40 years. In loop quantum gravity the discrete structure of the theory at the Planck scale suggests that a given (macroscopic) ADM mass configuration need not correspond to a unique quantum state. High degeneracy due to the contribution of microscopic degrees of freedom is expected but hard to prove at the present stage. This leads to a certain degree of disagreement on the status of such statements in the field at large. Although, a key instance where such degeneracy is accepted with little controversy is in the loop quantum gravity models designed to calculate black hole entropy (for reviews and references see \cite{BarberoG:2015xcq, Perez:2017cmj}) where the statistical origin of the entropy lies precisely in the large multiplicity of underlying microscopic states. Our simple model might still be too simple to represent definite evidence in this direction. Nevertheless, the results of Section \ref{degeM} do provide a toy model to eventually study the implications of the large degeneracy of the mass spectrum in discussion of the fate of information in black hole evaporation.

The model we introduce here is simple and workable, we hope it could provide potentially useful insights in dealing with qualitative questions concerning black hole evaporation. The investigation of these interesting possibilities is left for the future.

\section{Acknowledgements}

We thank the interaction with Simone Speziale for valuable insights and specially for discussion on the hamiltonian analysis of the system.

\providecommand{\href}[2]{#2}\begingroup\raggedright\endgroup

\end{document}